\documentstyle[psfig,prl,aps]{revtex}
\begin{document}
\draft
\twocolumn[\hsize\textwidth\columnwidth\hsize\csname@twocolumnfalse\endcsname
\title{Quantum and Classical Orientational Ordering in Solid Hydrogen}
\author{I. I. Mazin, Russell J. Hemley, A. F. Goncharov,
Michael Hanfland, and Ho-kwang Mao}
\address{Geophysical Laboratory and Center for High-Pressure Research\\
Carnegie Institution of Washington,
Washington, D.C.  20005-1305 }

\maketitle

\begin{abstract}
We present a unified view of  orientational ordering in phases
I, II, and III of solid hydrogen.  Phases II and III are
orientationally ordered, but the ordering objects in phase II are angular
momenta of rotating molecules, whereas in phase III 
the molecules order themselves.
This concept provides quantitative explanation for the vibron softening,
libron and roton spectra, and the increase of the vibron effective charge
in phase III, as well as a framework for understanding the topology
of the phase diagram and ortho-para state at high pressure.
The effective charge and the IR and Raman vibron frequency shifts are all
linear in the order parameter in phase III.
\end{abstract}

\pacs{} ]

\begin{figure}[tbp]
\centerline{\psfig{file=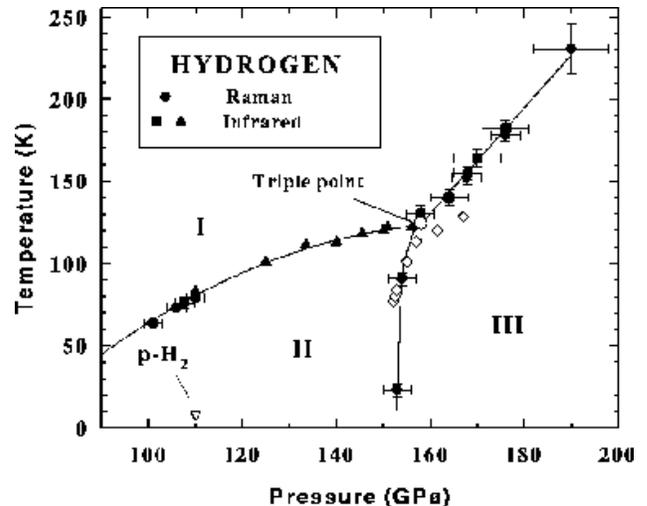,width=0.95\linewidth}}
\caption{Phase diagram of hydrogen at megabar pressures obtained from Raman
and infrared measurements \protect\cite{1,9,Oji}.  The open triangle 
and diamonds are from Ref. \protect\cite{lor}} \label{diag} \end{figure}

In the last decade, a wealth of information about the phase diagram of
hydrogen at high pressures has been collected.  However, even a qualitative
understanding of the phase transformations in the solid is  not yet
in hand.  Three phases are known in the experimentally accessible
range of pressures, which extend to 300 GPa and span over an order of
magnitude in compression (Fig. \ref{diag}).
The high-temperature phase consists of a closed packed lattice of freely
rotating and, on average, spherically symmetric molecules \cite{1}. The two
low-temperature phases have lower symmetry,
which implies at least partial ordering of the molecules\cite{1,9,lor}.
The properties of the two phases are so distinct, however, that more general
concepts than mere crystallographic dissimilarities must be invoked to
understand the origin of various phenomena they exhibit.  
The most unusual features include a
dramatic increase in the infrared activity of the main vibron, 
a strong vibron softening, an unconventional geometry of 
the phase boundaries, and drastic changes in rotational 
excitations\cite{9,lor,2}.  Here we show that these
disparate observations can be understood in terms of the concept 
of quantum versus classical orientational ordering in the dense solid.

Hydrogen at low temperature and pressure 
forms the only molecular quantum
solid.  As such, the excitations of the 
freely rotating molecules in phase I 
can be described by the rotational quantum number $J$.  This
contrasts with heavier molecular crystals,
where molecular rotation is substantially hindered even at low
pressures\cite{kran}. The properties of H$_2$ and D$_2$ crystals
greatly depend on the ground state of the constituent molecules ($J$=0 or 1),
which in turn is determined by their total nuclear spin.
Here we consider hydrogen solids having a substantial
fraction of spherically symmetric ($J$=0)
molecules (para-H$_2$ or ortho-D$_2$). Solids containing a sufficiently
high concentration of $J$=1 molecules (e.g., $o$-H$_2$/$p$-D$_2$) transform
to a cubic phase (space group {\it Pa}\={3}) even at ambient pressures 
(e.g., 3.8 K) \cite{kran}.
In this transition, the direction of angular momenta of individual molecules
orientationally order, a transformation driven by quadrupolar
interactions, where the ordered phase is the lowest-energy configuration
for classical quadrupoles\cite{kran}.

A different kind of orientational ordering occurs in phase II.  
In this case, solids consisting of spherically symmetric molecules 
($p$-H$_2$ or $o$-D$_2$) transform to ordered (broken symmetry) phases 
at $\sim$110 GPa in $p$-H$_2$\cite{lor} 
or at $\sim$28 GPa in $o$-D$_2$\cite{Wijn}
at $T\rightarrow 0$ K (Fig.\ref{diag}).  This transition was interpreted as
arising from increasing intermolecular interaction that results in mixing of
higher angular momenta (e.g., $J=\,2,\,4$ for $p$-H$_2$) into
the ground state molecular wave function and imparting a finite angular
moment for the molecules \cite{Wijn}.  The non-zero angular momenta can
then order, with molecular centers forming a lattice derived from hcp
\cite{1}.  Even more exotic ordering
schemes may exist for normal H$_2$ or D$_2$ \cite{Alex}.  We will refer
to these low-pressure ordered phases collectively as  phase II.
Here, ortho-para distinctions are valid in
the sense that the wave function can be defined for individual 
molecules and is either even or odd.

One expects that at still higher pressure intermolecular interactions
become so strong that all higher $J$ states have the same weight; that is,
the molecules behave classically and can orientational order as classical
rotors.  Such a pressured-driven transformation between two kinds of
orientationally ordered states would be a transition between a quantum and
classical crystal. This would be unique in relation to known quantum
crystalline transitions (e.g., solid He) in the sense
the quantum/classical distinction considered here is associated with
rotational degrees of freedom. As discussed above, 
spectroscopic changes at the II-III transition near 150 GPa
are significantly larger than those due to orientational ordering 
at the I-II boundary.
Not surprisingly, the higher pressure transition has been the focus
of considerable theoretical study recently (e.g., \cite{3,A,Tse}).
We show that new and previously reported data, including the
orientational order parameter, vibron effective charge, changes in
roton and libron spectra, geometry of the phase diagram, and
evolution of the ortho-para state, point to
the transition to phase III as being such a transformation. This
provides a simple and transparent model for orientational ordering in
phases I, II, and III.

\begin{figure}[tbp]
\centerline{\psfig{file=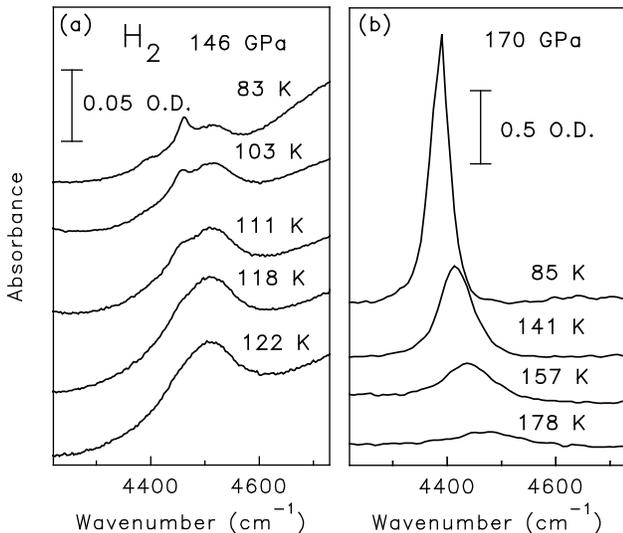,width=0.95\linewidth}}
\caption{Temperature dependence of IR vibron spectra. (a) I-II transition.
(b) I-III transition.}
\label{Fig1}
\end{figure}

{\it 1. Order Parameters and Infrared Intensity.}
We first consider the increase in IR vibron absorption associated
with passage into phases II and III (Fig. \ref{Fig1}).  At 167 GPa and 85 K,
the effective charge of the vibron is $q^{*}\approx 0.032e$ \cite{sci}; it
increases with pressure and reaches 0.037$e$ at 230 GPa. 
In phase II, in contrast,  $q^{*}$ is $\alt 0.004e$ at 140 GPa and the same
temperature, whereas phase I is characterized

\begin{figure}[tbp]
\centerline{\psfig{file=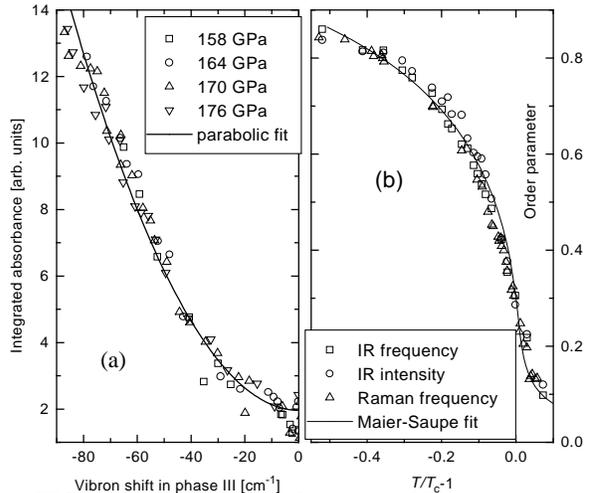,width=0.95\linewidth}}
\caption{(a) Experimental integrated IR intensity versus frequency shift
$\Delta \nu $ for phase III. 
(b) Order parameter from IR intensity, and IR and Raman
frequency shifts at 176 GPa.} \label{Fig3} \end{figure}

by a broad, weak disorder-induced band \cite{2,Oji,19}. 
The origin of these differences
can be understood from analysis of the temperature dependence of
$q^{*}$. We find a striking correlation 
between the temperature dependence of
$(q^{*})^2$ (integrated intensity) and $\Delta \nu $, the frequency shift
with respect to its value at the transition point ($P_c,T_c$)
for phase III; i.e., $(q^{*})^2\propto \Delta \nu^2$ 
over the entire {\it P-T} range investigated (Fig. \ref{Fig3}a).
The observation of $q^{*} \propto \Delta \nu$
can be interpreted to mean that both quantities are linear
functions of a scalar order parameter, $\Delta \nu \propto \eta $, and
$q^{*}\propto \eta $ \cite{isnote}. Recent measurements of the Raman vibron
in phase III (for D$_2$)\cite{13} showed that the temperature dependence of
the frequency shift (proportional to $\eta$) can be described by a
Maier-Saupe model, which characterizes the orientational ordering of
classical rotors and initially was derived for liquid crystals.  We show
here that the same is true for the IR intensity and both IR and Raman
shifts \cite{1,lor} (Fig.\ref{Fig3}b). In contrast, the normalized order
parameter in phase II is qualitatively different, being
much steeper as a function of $T/T_c$ ({\it cf.,} Ref. \cite{19}).

{\it 2. Magnitude of the Effective Charge.}
We now show that not only the temperature dependence of $\eta$, but also
the magnitude of $q^{*}$ can be understood within the
proposed framework. Previously, the origin of
the vibron intensity in phase III was examined from various electronic
standpoints\cite{3,A,Tse}.  We note here that a condition for
vibron IR activity is that the two atoms in a molecule are
crystallographically 
inequivalent \cite{zall}.
If the molecules have non-zero quadrupole moments, 
a non-zero electric field is then induced at the
lattice sites, which in turn polarizes the  molecules and creates an
effective dipole moment\cite{clarif}.  We have  estimated the magnitude of the
effect for H$_2$ by calculating the induced field in several well-studied
quadrupolar lattices.  The total field is proportional to the quadrupole
moment, $E=\beta Q/a^4$, where $a$ is the lattice parameter.
The component of the field along the H-H bond,
$E_{\parallel }$, is $\beta _{\parallel }Q/a^4$. The values for
$\beta $ depend on the structure and orientation, but for a
large class of structures they are of order unity \cite{beta}. 
For a static molecule in the
zero-pressure solid, $Q\approx $ 0.5 a.u.\cite{kran} and near the II-III
transition $a$ is $\sim $3.3 bohr. At the equilibrium bond length 
($d=1.5 $ bohr) the polarizability $\alpha _{\parallel }$ of
the H$_2$ molecule is
6.72 bohr$^{-3}$ \cite{Kolos}, which gives an effective
static charge $q=\alpha E/d\approx 0.025e$. To find the dynamic charge,
we substitute $\alpha $ by $d(\partial \alpha /\partial d)$. From Ref. \cite
{Kolos}, we have $q_d\approx 1.5q$, which gives $q^{*}\approx 0.06e$ for the
total effective charge, close to the experimental value of 
0.03-0.04$e$ \cite{order}

This result should be contrasted with the behavior of the vibron in
``quantum'' phase II. In this phase, the narrow absorption peak appears on
the shoulder of the broad disorder-induced band (Fig. \ref{Fig1})\cite
{1,2}. This sharp peak 
is a symmetry-allowed IR-active vibron and
signals the onset of orientational ordering of angular
momenta. The maximum intensity of this vibron 
($q^{*}\approx 0.004e$) is an order of magnitude smaller than that for
phase III (e.g., at $165$ GPa) at the same temperature\cite{2,sci}.
Several factors reduce the effective charge in phase II compared to that in
III: (a) In a pure quantum state in which
$\{J,M_J\}=\{l,l\}$, the expectation value
of {\it Q }is reduced by a factor
of $(2l+3)/l$\cite{J1} with respect to that of a static molecule,
and (b) the polarizability perpendicular to the
molecular bonds is smaller than that parallel.
The smaller polarizability not only diminishes the static charge by 40\%,
but also reduces the dynamic charge by a factor
of three. As a result, the effective charge is reduced by 2--3.5 from the
change in $\langle Q\rangle $ and 2--2.5 from the polarizability change,
giving a total reduction by a factor of 5-10, in agreement with experiment.

\begin{figure}[tbp]
\centerline{\psfig{file=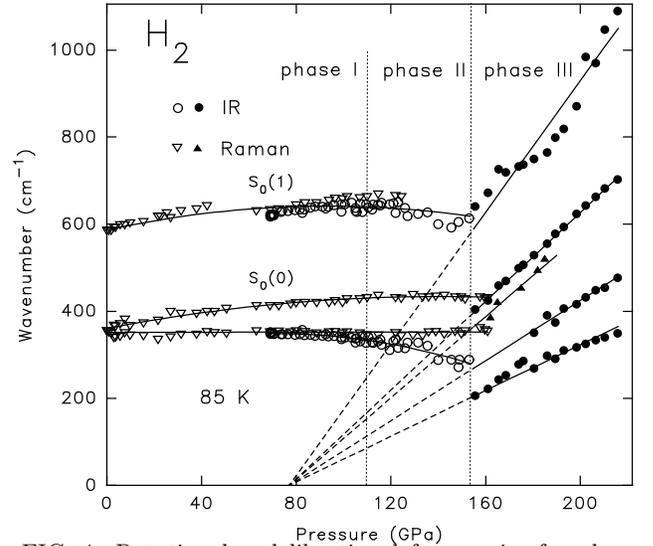,width=0.95\linewidth}}
\caption{Rotational and librational frequencies for phases I, II and III of
hydrogen (85 K)\protect\cite{1,2,sci}.  The IR
data were obtained by subtracting the vibron frequency
from IR combination bands measured.  A new
Raman band in phase III having a strong pressure
dependence identical to one of the IR modes is shown.  The dashed line
suggests ``latent soft mode'' behavior of the librons. }
\label{rotons}
\end{figure}

{\it 3. Rotons and Librons.}
The proposed picture further implies that elementary excitations
corresponding to the angular degrees of freedom in the system (rotons and
librons) must be very different in phases II and III. Indeed, IR and Raman
data reveal a striking change in these low-frequency excitations upon
crossing the II-III phase boundary. Fig.\ref{rotons} shows rotational and
librational mode frequencies from new and earlier spectroscopic
data up to 230 GPa.  No discontinuity 
is observed in the broad roton bands at the I-II transition 
for normal H$_2$ (110 GPa and 85 K).  
In contrast, the rotons disappear at
the II-III transition, and are replaced by new excitations in
the same energy range.
This change can be understood by considering the difference in libron
spectra for quantum and classically ordered phases. For the former, $J$
remains a good quantum number and the roton excitations characteristic of
the disordered phase are supplemented by $M_J\rightarrow
M_{J^{\prime }}$ ``libron'' excitations of the ordered phase, as in the
{\it Pa}\={3} structure of ortho-rich H$_2$ \cite{kran}. Likewise, phase II 
consists of {\it rotating} molecules with a similar set of excitations. In
contrast, the molecules in phase III behave as classical objects (as in
solid N$_2$), and $J$ is no longer a good quantum number. The elementary
excitations associated with rotational degrees of freedom are neither
$J\rightarrow J^{\prime }$ rotons nor $M_J\rightarrow M_{J^{\prime }}$
transitions but are anharmonic classical librons, quantized angular
oscillations about the equilibrium orientation in which full rotation is
strongly hindered. Moreover, the phase III librons harden
dramatically with pressure and extrapolate to zero about 70
GPa below transition (Fig.\ref{rotons}).  This strong pressure
dependence is expected for librons and can be contrasted with the behavior
of the higher frequency lattice mode, which shows at most a small
discontinuity across the II-III transition \cite{1}.
These observations, together with the continuous and discontinuous changes
in the vibrons\cite{13}, are indicative of a first-order, but
non-reconstructive,
mechanism for the transition.

{\it 4. Geometry of the Phase Diagram.}
The geometry of the phase boundaries near the I-II-III triple
point (Fig.\ref{diag}) also fits the picture of a ``quantum-classical''
orientational transition. Near the triple point, now established for H$_2$
and D$_2$ \cite{1,Oji,19,13},
the I-II boundary is nearly horizontal, $dT_c/dP_c\alt%
0.4$ K/GPa, while the II-III boundary is nearly vertical
up  to $\sim$80 K, $dT_c/dP_c
\agt30$ K/GPa in H$_2$;
this effect is even more pronounced in D$_2$ \cite{13}.  Such a
configuration is highly unusual for a crystallographic transition.  One can
use the Clapeyron equation, $\Delta V/\Delta
S=dT_c/dP_c,$ to estimate the volume change across each phase transition.
Assuming an entropy change $Delta S$ $\approx$ 0.2$R$/mole 
(Ref.\cite{kran}, p. 238), we estimate $Delta V$ 
at the I-II transition to be 0.0007 cm$^3$/mol.  
An analogous calculation for the I,II-III transition gives 
about 0.05 cm$^3$/mol. 
Indeed, one should expect large changes in intermolecular forces 
for the latter because of modifications in 
quadrupole-quadrupole interactions \cite{1}.  This effect
can be estimated for static molecules from the 
electrostatic energy of quadrupolar closed-packed lattices, $U=1.5\kappa
Q^2/a^5$, $\kappa \approx 7$ \cite{kran}.  Differentiating this with
respect to volume, we calculate an ``electrostatic'' pressure
of about 8 GPa, which gives 0.05 cm$^3$/mol at 150 GPa\cite{EOS}.
Note that the quadrupolar energy is reduced by a 
factor of 4/25\cite{kran} for rotating $J=1$ molecules,
which is in qualitative agreement with
the much smaller volume change at the I-II transition.

{\it 5. Ortho-Para State.}  The ``quantum-classical'' transition associated
with phase III has important implications for the evolution of
ortho-para states as a function of pressure.
There is evidence for ortho-para distinguishability under
moderate pressures, but its persistence to the megabar range has been a 
subject of debate \cite{1}. The above
analysis naturally leads to the conclusion that the wave function of the
crystal in phase III can no longer be factored into individual molecular
wave functions, so single-molecule ortho-para
distinguishability is lost.
In contrast, despite the mixing of $J$ states in phase II, 
the notion of parity of the wave function of nuclei in an
individual molecule is valid.  This is supported by recent measurements for
D$_2$: in phases I and II significant differences in
IR and Raman bands are observed for samples starting out as pure
ortho versus normal\cite{Alex}; in phase III, however,
the number and frequencies are identical and independent of
sample history \cite{Oji,19}.

In conclusion, we suggest that orientational ordering in
the higher pressure molecular phase of hydrogen (phase III) 
is qualitatively different from that observed at 
lower pressures (phases I and II).  
The molecules in phase III are orientationally ordered in the
sense that the time average of the molecular bond direction
is non-zero, whereas in phase II the ordering objects are 
axes of quantization of angular momentum.
This concept provides a natural explanation of the striking
IR vibron activity and other properties of  phase III. Further quantitative
treatment should take into account detailed
structural changes occurring at these transitions as well as the effects
of zero point motion (e.g., Ref.\cite{A}).

We thank M. Li and J. F. Shu for assistance, and N. W. Ashcroft and
J. H. Eggert for comments. This work was supported by the NSF and NASA.


\begin{references}
\bibitem{order} The calculated value is expected to be larger because
the effects of finite temperature and zero-point motion are
neglected in the model.
\bibitem{J1}  If $J>1$, the state with $M_J=0$ has the largest absolute
value of the quadrupolar moment (which is negative); for $J=1$, however, the
$M_J=0$ state has twice the expectation value of {\it Q} relative to $%
M_J=\pm 1$ (and is positive).
\end{references}
\end{document}